\documentclass[epj]{svjour}

\usepackage{graphicx}
\usepackage{fancyhdr}

\def\mytitle{My title} 
\def\myauthors{My name}  
\def\mytype{My type of session}
\def\mysession{My session}

\def\mytitle{Short title of talk} 
\def\myauthors{Name of Author}    
\def\mytype{Contributed Talk}    
\def\mysession{Flavor Physics}

\begin{document}
\title{Neutrino time travel}
\author{James Dent\inst{1}
\and Heinrich P\"as\inst{2}
\and Sandip Pakvasa\inst{3}
\and Thomas J. Weiler\inst{1}
}                     
%
%
\institute{Department of Physics and Astronomy, 
Vanderbilt University, Nashville, TN 37235, USA
\and 
Institut f\"ur Physik, Universit\"at Dortmund, D-44221 Dortmund, Germany
\and
Department of Physics \& Astronomy, 
University of Hawaii at Manoa,
2505 Correa Road, Honolulu, HI 96822, USA
}
%
\date{}
\abstract{
We discuss causality properties of extra-dimensional
theories allowing for effectively superluminal bulk shortcuts. 
Such shortcuts for sterile neutrinos have been discussed as a 
solution to the puzzling LSND and MiniBooNE neutrino oscillation results.
We focus here on the sub-category of
asymmetrically warped brane spacetimes
and argue that scenarios with two extra dimensions
may allow for timelike curves which can be
closed via paths in the extra-dimensional bulk. 
In principle sterile neutrinos propagating
in the extra dimension may be manipulated in a way 
to test the chronology protection conjecture experimentally.
\PACS{
      {04.20.Gz}{Spacetime topology, causal structure}
\and  {14.60.St}{String and brane phenomenology}
\and  {11.25.Wx}{Sterile neutrinos}
     } 
} 
\maketitle
\section{Introduction}
\label{intro}

In the Marvel superhero
comics, the time keepers are unfriendly, greenish looking guys with
insect faces, which 

{\it ``were born at the end of time, entrusted 
with the safety of 
Time by He Who Remains. The Time Keepers were meant to watch over the space 
time continuum just outside of Limbo, and make sure the universe thrived.''}
\cite{timekeep}

On planet Earth, the representative of the time keepers is Stephen W. Hawking,
who announced in his chronology protection conjecture that

{\it ``It seems that there is a chronology protection agency which 
prevents the
appearance of closed time-like curves and so makes the universe safe for 
historians''} \cite{hawking}. 

Hawking had good reasons for this statement: the ideas that
exist in the literature for spacetimes which allow for closed timelike
curves (CTCs), including wormholes, Kerr- or Kerr-Newman black holes and
Tipler cylinders (for an overview see \cite{Visser:2002ua}), 
typically are found to
suffer from one or more of the following obstacles:

\begin{itemize}

\item
negative energy densities violating the so-called energy conditions
may be required to warp spacetime into the 
desired geometry. Such negative energy seems to be unstable
to small perturbations, at least in semi-classical 
calculations (for a recent discussion see
\cite{buniy}).

\item
quantum corrections to the stress-energy tensor seem to diverge close to
the chronology horizon which separates spacetime regions allowing for CTCs
from the regions in which they are forbidden, 
again at least semi-classically 
 \cite{hawking}.

\item
nobody has ever spotted a wormhole or the region inside the black hole
horizon, and Tipler cylinders have to be unreasonably large.

\end{itemize}

While these facts do not constitute a proof that the corresponding spacetimes
or time travel in general 
are impossible -- after all no complete theory of quantum
gravity
which could decide the issue exists -- 
they make these possibilities unlikely in the view of most physicists. 

Here we discuss the example of an extra-dimensional brane universe with CTCs, 
which alleviates at least some of
these problems. Moreover, what is particular attractive:
this scenario is in principle 
experimentally testable in neutrino experiments.

\begin{figure*}[!t]
\centering
\includegraphics[clip,scale=0.50]{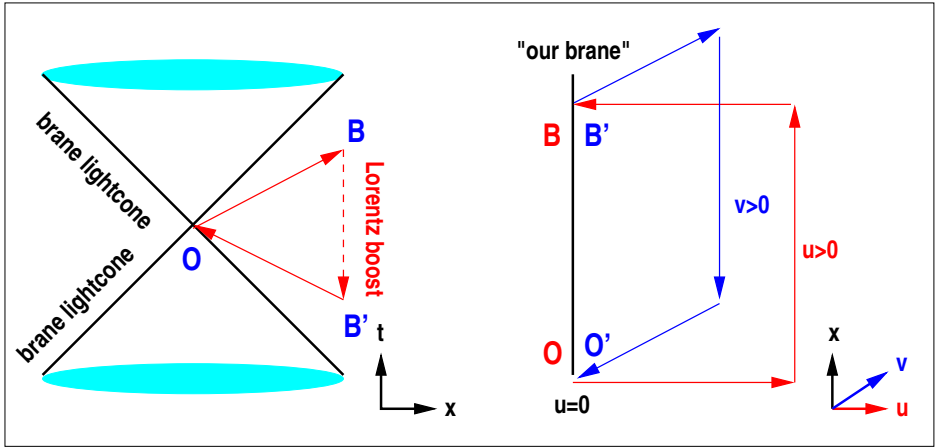}
\caption{Closed timelike curve in an asymmetrically warped universe:
(i) A signal takes a space-like shortcut via a path of constant
$u=u_1$ with $v=0$ from point $O$ to point $B$.
(ii) A Lorentz boost transforms $B$ into $B'$ with negative time coordinate. 
(iii) A return shortcut at constant 
$v=v_1$ with $u=0$ closes the timelike curve.
}
 \label{bulkpath}
\end{figure*}

\section{Asymmetrically warped brane universes and CTCs}

In his SUSY'07 presentation one of us (HP) discussed the possibility, that the
LSND \cite{lsnd} neutrino oscillation anomaly and the MiniBooNE 
null result \cite{miniboone}
can be explained
by oscillations into a sterile neutrino taking
shortcuts in extra dimension
\cite{ppw}. This scenario also potentially fits \cite{ppwplus} the
anomalous 
resonance-like structure seen in the low-energy bins of the MiniBooNE data 
\cite{miniboone}.
All simple scenarios of this type are causally 
stable.
This includes scenarios where the shortcut is realized by asymmetrical warping
of an extra dimension, described by a spacetime metric of the type
\cite{freese,erlich} \footnote{Such spacetimes have also been discussed as a
  solution to the cosmological horizon and dark energy problems.}
\begin{equation}\label{metric}
ds^2 = dt^2 - \sum_i \alpha^2(u)\,(dx^i)^2 - du^2,
\end{equation}
with a warp factor $\alpha^2(u)<1$ allowing for effectively superluminal
propagation in the extra dimension. 
Here our brane is located at the $u=0$ sub-manifold and
the index $i$ runs over the spatial dimensions $i=1,2,3$ parallel to the brane.
In this note, however, we keep the promise made as a concluding remark
in the SUSY'07 talk:
``if you are
desperate to have a neutrino time machine I'll get you one.''

In order to do so we consider two asymmetrically warped extra dimensions
 ``$u$'' and ``$v$'' with warp factors $\alpha(u)$ and
$\eta(v)$, respectively
\cite{ctc}. We assume that the $u$- and $v$ dimensions assume the 
simple form (\ref{metric}) in different Lorentz frames. Note that this
is natural for any spacetime with two or more extra dimensions, as there is
no preferred Lorentz frame from the viewpoint of the brane. In the following we
construct this 6-dimensional metric explicitly. We denote the relative velocity
between the two Lorentz frames, in which the $u$ and $v$ dimensions assume the
simple form (\ref{metric}) as $\beta_{uv}$.

It is easy to show, then, that the full 6-dimensional metric assumes the form
\begin{eqnarray}
\label{6dmetric}
ds^2&=&
 \gamma^2 
\left[
(1-\beta_{uv}^2 \eta^2(v)) dt^2
 - 2 \beta_{uv} \alpha(u) (\eta^2(v)-1) dx dt 
\right. 
\nonumber
\\ &&
\left.
-\alpha^2(u) (\eta^2(v) - \beta^2_{uv}) dx^2
\right] 
- du^2 - dv^2.
\end{eqnarray}
For $v=0$ (\ref{6dmetric}) reduces to 
(\ref{metric}),
for $u=0$ (\ref{6dmetric}) reduces to (\ref{metric})
boosted by 
$\beta_{uv}$,
and for $u=v=0$, 
i.e. on the
brane, to 4-dimensional Minkowski spacetime. By boosting the metric 
(\ref{6dmetric}) with 
$\beta=-\beta_{uv}$ the $v$-dimension assumes the simple form (\ref{metric})
and the metric for the $u$-dimension becomes non-diagonal.

If the $x$ coordinate would be periodic, the $u=0$ slice of the metric 
(\ref{6dmetric}) would allow for a simple mapping into
the Tipler-van-Stockum spacetime, which is well known to accommodate
CTCs \cite{tipler}.
However, in the case of a boosted 
asymmetrically warped extra dimension, the variable $x$ is not periodic
(unless our universe has the topology of a flat torus).
It is thus required to construct an explicit return path to the spacetime
point of origin, to close the CTC. 

\begin{figure*}[!t]
\centering
\includegraphics[clip,scale=0.7]{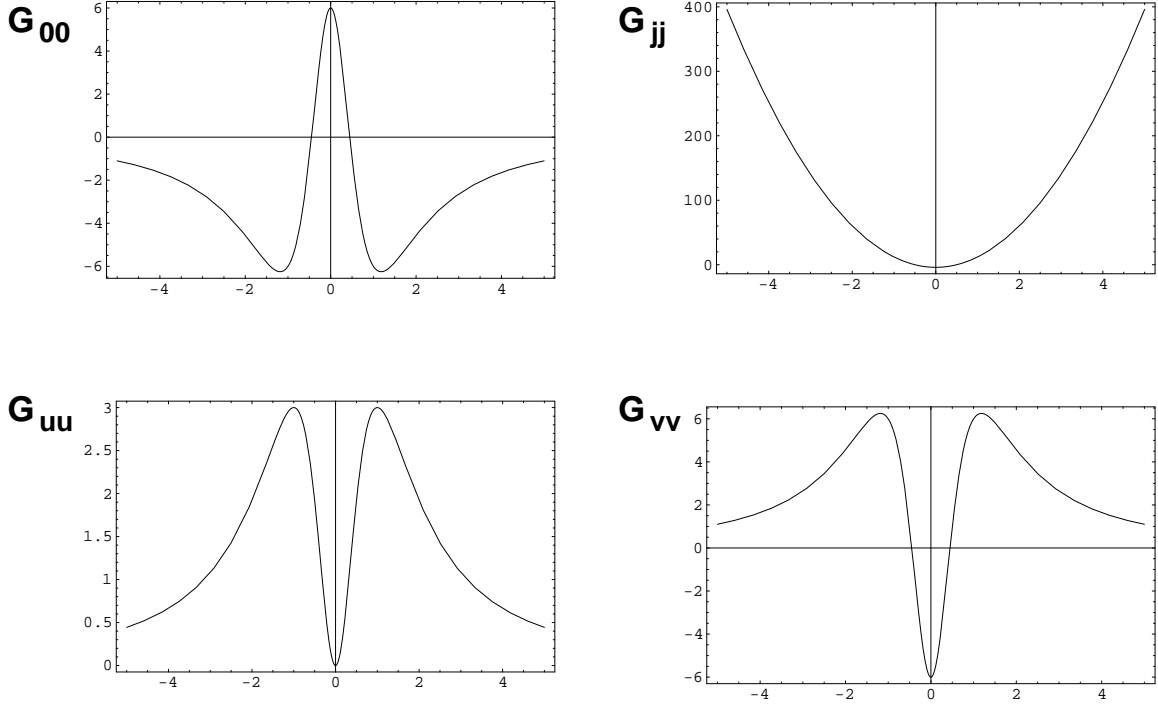}
\caption{
Nonzero elements of the Einstein tensor $G^{\mu\nu}$
on the $v=0$ slice, as a function of $u$, for
warp factors $\alpha(u)=1/(u^2 + c^2)$ and $\eta(v)=1/(v^4 + c^2)$, 
with $c=1$. While the null, weak and dominant energy
conditions are violated in the bulk, they are conserved on the brane.
}
 \label{encond}
\end{figure*}

In the following we consider a signal following a particular 
path as given in Fig.~\ref{bulkpath}. The signal leaves 
our brane at the spacetime point $O=(t=0,x=0, u=0)$, 
and propagates on the hypersurface at $u_1$ for a travel time $t$ with the
limiting velocity $(\alpha(u_1))^{-1}\equiv\alpha_1^{-1}$.
We will assume that $0 < \alpha_1 <1$, so that the travel speed 
in the bulk 
is superluminal relative to travel speed on our metric. 
At later time $t$, 
the signal may 
reenter our brane.
In the limit $u_1\ll \alpha_1^{-1}t$,
which is always fulfilled for sufficiently large $t$, 
the reentry point on our brane is $B^\mu\approx (t,x=\alpha_1^{-1}t, u=0)$.
Since the distance to the reentry point $B^\mu$ is space-like
(i.e. outside the brane's lightcone), 
it may be transformed to negative time by a boost on our brane.  
The boosted point $B'^{\mu}$ has coordinates
\begin{equation}\label{ltrans2}
x'=\gamma\,t\left(\alpha_1^{-1}-\beta\right),~~~~
t'=\gamma\,t\left(1-\beta \alpha_1^{-1}\right).
\end{equation}
It is clear that for
\begin{equation}
0<\alpha_1<\beta<1
\end{equation}
an observer in the boosted frame on our brane sees the signal arrive  
in time with $t'<0$, i.e., before it was emitted.  
However,
this result alone does not imply any conflict with causality.
In particular, it does not necessarily imply that the 
spacetime is blessed with 
CTCs. 
To close the timelike curve, 
one has to show that the time $t'$ during which 
the signal traveled backwards in time, 
is sufficiently large to allow a return from the
spacetime point $B'^{\mu}=(t',(x=\alpha_1^{-1}t)',0)$ to the 
spacetime point of origin, $O=O'=(0,0,0)$. 
The speed required to close the lightlike curve of the signal,
as seen by the boosted observer on the brane, is 
\begin{equation}
\label{creq}
c'_{\rm req}=\frac{(x=\alpha_1^{-1}t)'}{|t'|}
=\frac{1-\beta \alpha_1}{\beta-\alpha_1}\,,
\end{equation}
where the latter expression results from using (\ref{ltrans2}).
It is easy to show that the condition 
$0<\alpha_1<\beta<1$ implies that $c'_{\rm req}$ itself 
is superluminal.
Thus there is no %
return path on our brane which leads %
to a CTC.  

We now benefit from the fact that the spacetime considered has
two extra dimensions and that we have
the possibility to use the
$v$ dimension to chose an alternative return path from the boosted point $B'$ 
to $O$, allowing for $|c'_{\rm bulk}|> |c'_{\rm req}|$.
For simplicity we choose $\beta=-\beta_{uv}$ and the return path in 
the $u=0$, $v>0$
plane, where the $x-v$ slice of the metric takes the simple form
(\ref{metric}).
Now
any suitably chosen 
warp factor $\eta(v)$ with
$\left(\eta(v_1)\right)^{-1}
\geq c'_{\rm req}$ in combination 
with $\beta_{uv}> \alpha(u_1)$ generates a CTC.

\section{Stress energy tensor, energy conditions and other
pathologies}

In view of the stability pathologies related to the requirement of
negative energy densities for spacetimes 
with CTCs known in the literature,
it is interesting
to analyze, whether the stress-energy tensor 
\begin{equation}
T_{\mu \nu}=\frac{1}{8\,\pi\,{\rm G_N}}\,G_{\mu \nu}
\end{equation}
for the extra-dimensional metric (\ref{6dmetric}) fulfills the 
energy conditions. The
null, weak,
strong and dominant energy 
conditions are defined as follows,
\begin{eqnarray}
{\rm NEC}&:& \rho + p^j \geq 0 ~~~\forall j; \\
{\rm WEC}&:& \rho \geq 0 ~~~{\rm and}~~~ \forall j,~~~  \rho + p^j \geq 0;\\
{\rm SEC}&:& \forall j,~~~ \rho + p^j \geq 0 ~~{\rm and}~~  
\rho + \Sigma_j p^j \geq 0;\\
{\rm DEC}&:& \rho \geq 0 ~~{\rm and}~~ \forall j,~~~  p^j \in [\rho,-\rho].
\end{eqnarray}
While it is in principle 
straightforward to calculate the 
Christoffel symbols 
from the metric (\ref{6dmetric}),
the general expressions 
 are complicated.
Thus we constrained ourselves to a numerical analysis 
of specific examples
for the warp factors $\alpha(u)$ and $\eta(v)$.

It is not difficult to find a functional form for the
warp factors $\alpha$ and $\eta$ , which conserves some of 
the energy conditions, at 
least on 
the brane. One such example is given by 
$\alpha(u)=1/(u^2 + c^2)$ and $\eta(v)=1/(v^4 + c^2)$. For this case 
the 
elements of the Einstein tensor on the $v=0$ slice
are shown as a function of $u$ in 
Fig.~\ref{encond}. The null, weak and dominant energy conditions 
are
conserved on the brane, while the strong energy condition is violated on the 
brane (see Fig.~\ref{encond}).

This property of good energy conditions on the brane 
favors the scenarios discussed over most previous examples
of spacetimes with CTCs. Moreover, one could argue that also instabilities due 
to quantum corrections are less likely to diverge in a spacetime of higher
dimension. Finally, due to the universal access to 
the extra dimension this scenario 
has the particular advantage that the ambiguity due to the lack of a full 
quantum gravity treatment of the problem can be resolved: it allows - at least
in principle - for an experimental test.

\section{Neutrino time machine}

As mentioned before, a natural and well motivated
candidate for a bulk fermion in string theory is a gauge singlet 
``sterile'' neutrino. In many particle theories, these sterile neutrinos
would mix with the Standard Model neutrinos, and will be generated by neutrino
oscillations when neutrinos propagate in space and time.
Sterile neutrinos thus provide an accessible probe for the causality
properties of extra dimensions. An experiment testing for such properties
of spacetime could effectively transform active neutrinos  
into sterile neutrinos by using the resonant conversion in an
increasing matter density known as Mikheev-Smirnov-Wolfenstein (MSW) 
\cite{smi,wolf} effect. 
A neutrino beam of suitably chosen energy could be generated in a beam-dump
experiment at one of the Earth's poles and pointed into the ground, traversing
a pit with a slowly changing matter density profile.
Once being
converted inside the Earth's core, the sterile neutrinos will avail shortcuts 
in extra dimension to extremize the action of the propagation path, and thus 
effectively propagate a spacelike distance. On their way out of the
Earth's interior the sterile neutrinos are
reconverted into active flavors, which could be observed by a neutrino
detector at the equator. If the neutrinos have advanced a spacelike
distance, the Earth's spin at the equator will transform the signal into
a moving reference frame and thus reverse the order of emission and detection.
It could be possible to send the signal back to the point of origin 
(the pole lab) before it had been sent off.

\section{Conclusions}

In this paper we discussed the causality properties of a class of
brane universes with
two asymmetrically warped extra dimensions. Such scenarios are attractive in 
view of the LSND and MiniBooNE anomalies, as well as due to their potential
to solve or alleviate the
cosmological horizon and dark energy problems \cite{freese,erlich}. 
We have shown that  
rather generic examples of such spacetimes have closed timelike curves without
violating the null, weak and dominant energy conditions
on the brane. Moreover, neutrino
oscillations may provide a unique possibility to test Hawking's chronology
protection conjecture, should such spacetimes really be realized in Nature.
A realistic description of this process, however,
would require a quantum field theoretic treatment 
similar to the one preformed in \cite{Burgess:2002tb}.
We conclude that neutrinos might not only serve as
a probe for physics beyond the
Standard model and cosmology, but also for the understanding of the 
deepest foundations of causality and time.

\section*{Acknowledgements}
JD and TJW were supported by the US Department of Energy under Grant
DE-FG05-85ER40226. HP was supported by the
the University of Alabama and the EU project ILIAS N6 WP1. 
SP was supported by the
US Department of Energy under Grant
DE-FG02-04ER41291.

\end{document}